# The Los Alamos Computing Facility during the Manhattan Project

B.J. Archer[1]
Los Alamos National Laboratory
Los Alamos, New Mexico 87545

**Abstract:** This article describes the history of the computing facility at Los Alamos during the Manhattan Project, 1944 to 1946. The hand computations are briefly discussed, but the focus is on the IBM Punch Card Accounting Machines (PCAM). During WWII the Los Alamos facility was one of most advanced PCAM facilities both in the machines and in the problems being solved.

## I. INTRODUCTION

Prior to World War II (WWII) there were three types of scientific computing machines. The most common were the desk calculators used by human computers to manually solve equations, see Figure 1. There were also a handful of analog differential analyzers. The most advanced differential analyzers were developed by Vannevar Bush at MIT during the late l930's and early 1940's.[1] These mechanical systems could solve second-order differential equations.[1]

The third computing method was to adapt punch card machines to solve scientific equations.[1] In 1928 Leslie Comrie at His Majesty's Nautical Almanac adapted punch card machines to calculate tide tables based on the orbit of the moon. In 1937 Wallace Eckert persuaded International Business Machines (IBM) to support the Watson Astronomical Computing Bureau at Columbia University in New York with IBM Punch Card Accounting Machines (PCAM). The Astronomical Bureau was dedicated to calculating celestial orbits. Eckert published a book on scientific punch card computing in 1940, "Punched Card Methods in Scientific Computing" that had wide influence.[1,2] In 1940 Eckert moved to the US Naval Observatory where he established a PCAM facility to calculate star navigation tables for aircraft and ships. In 1941 a PCAM facility was established at the Army Ballistics Research Laboratory (BRL) at Aberdeen Maryland to calculate artillery ballistic tables.[1]

At Berkeley in the summer of 1942 Robert Oppenheimer and Robert Serber had Stanley Frankel and Eldred Nelson work out the neutronics integral theory to calculate the critical mass of a uranium bomb.[3,4,5] This work was pivotal at the June 1942 Berkeley conference that concluded a nuclear bomb was feasible.[4,5,6] As a result, Frankel and Nelson were early staff members at Los Alamos. In retrospect, this was a crucial point in developing the computing capability at Los Alamos.

A computing facility encompasses much more than just the computers. It includes the actual facility that houses the machines, the staff to operate the machines, the numeric methods for solving the equations, and the staff to setup and interpret the solutions. Los Alamos had all of this in place by April 1944 for the PCAM computing facility. During WWII the Los Alamos facility was one of the most advanced PCAM facilities, both in the type of machines and in the breadth and complexity of problems being solved.

This paper starts with the desk calculators. The next section describes the IBM PCAM. Next is a brief description of the uses of the PCAM. Late war developments are discussed next. The last section describes some of the post-war developments.

---

[1] Corresponding author: barcher@lanl.gov

## II. THE MECHANICAL CALCULATORS

Frankel and Nelson, Figures 2 and 3, had set up a "hand-computing activity to support development of the electromagnetic isotope separator and early critical-mass calculations" at Lawrence's Berkeley lab.[5, 7] Frankel and Nelson arrived at Los Alamos in early 1943 to continue the critical mass work for the gun bombs. They were tasked to order mechanical calculators for Los Alamos, they chose a variety of Marchants, Fridens, and Monroes.[7]

Some of these calculators were distributed to scientists, but others were used to set up the central computing pool, Group T-5, in the summer of 1943 that was led by Donald (Moll) Flanders,[7, 8] Figure 4. The original computers were the wives of the technical staff, who were eventually replaced by Women's Army Corps (WAC) staff.[9] Mary Frankel, Figure 5, set up the problems for the computers to calculate.[9] A trivial example could be linear interpolation of a table, $c = (a+b)/2$. Frankel would set up a calculating sheet with a row telling where to get "a" and a space for its value, a row to get "b", a row that says do "a+b" with a space for the value, and finally a row to divide by 2, with a space for the final value. This level of detail was needed for every operation, with complicated operations handed off between different computers.

Flanders decided to keep only the Marchant mechanical calculators, the most powerful ones,[7] which were used extensively into the early 1950s for smaller numerical calculations. These mechanical calculators were famously repaired by Feynman and Metropolis.[8, 9] T-5 continued through all the reorganizations of T-Division during the war.[6] See the companion ANS article for further details of the mechanical computing.[10]

## III. THE IBM PUNCH CARD ACCOUNTING MACHINES

In 1943 Los Alamos was focused on the gun bomb approach. A particularly vexing issue with the gun bomb designs was calculating the critical mass of "odd-shaped bodies" resulting from firing the gun.[4, 11] In the fall of 1943, it was clear that the hand calculations were struggling with this problem, and Dana Mitchell, Figure 6, recommended to Hans Bethe that the IBM Punch Card Accounting Machines (PCAM) should be used for the critical mass calculations.[5, 7, 11] Mitchell had worked at Wallace J. Eckert's laboratory at Columbia University, which had pioneered the use of the PCAM for astronomy calculations.[5, 7, 11] Bethe had seen Eckert's computing facility and decided to order PCAM for Los Alamos.[11]

Frankel and Nelson were assigned responsibility for determining how to use the IBM machines for the odd-shaped critical mass problem, and specifying the machines to order.[4, 5, 7, 8] In January 1944 Frankel visited the East Coast to get information about the suitability of the IBM machines for the odd-shaped criticality problem. Presumably Frankel visited the Astronomical Bureau at Columbia University in New York, the Nautical Almanac at the Naval Observatory, Washington D.C., and the Army Ballistics Research Laboratory (BRL) at Aberdeen Maryland. In January 1944 Frankel recommended proceeding with the PCAM. These were to be similar to those in use by the BRL for calculating ballistics tables[1] and the Nautical Almanac for calculating navigation tables.[12]

The IBM PCAM were delivered on April 4, 1944 (Ref. 4). Because of the classified nature of Los Alamos, IBM was not told where the machines were going and could not send a maintenance man to assemble them. However, IBM identified their best PCAM maintenance man who had been drafted, John Johnston, who was then assigned to Los Alamos by the Army.[7] The PCAM arrived at Los Alamos three days before Johnston, so Frankel, Nelson, and Richard Feynman assembled the system, and Johnston carried out the final tuning.[5, 7, 9, 11] The PCAM were installed in Building E, Figure 7 (Ref. 8).

There is some uncertainty about all the components of the PCAM system used during the war, but the eight core machines are agreed to by all sources.[7, 8, 13] These eight machines are shown in the first part of Table I. They were operated as a combined system to carry out a calculation. This was one of the largest PCAM installations in existence during the war.[1, 2]

A very detailed description of the IBM PCAM and the operation of the system was written in 1949 (Ref. 13). Figure 8 shows part of the Manhattan Project PCAM facility after the 601s with a divide capability, discussed later, were delivered in November 1944. Note that Figure 8 shows four 601 multipliers, so at least one of the original 601s was kept after the special 601s were delivered.

Figure 9 shows the Los Alamos PCAM facility after the first 602 was delivered sometime in 1947. It has expanded some, with three 077 collators, two 513 reproducing punches, and one of the new IBM 602 calculating punch machines. Otherwise, except for missing a 031 duplicating keypunch, this is the equipment used during the Manhattan project.

There are two auxiliary machines in Table I that only appear on single lists. One was a verifier of unknown model number that Nelson lists.[7] On a verifier an already typed card was typed again to check that it was typed in correctly. The 1949 list[13] has a 552 interpreter. This device was introduced in 1946, but was preceded by the 550 interpreter. The interpreter duplicated a card with human readable annotation across the top. As a practical matter, Los Alamos must have had an interpreter. Another bit of evidence for the existence of these two machines is that Bethe mentioned that he ordered 10 machines, not 8 (Ref. 4).

**IV. USING THE PCAM**

John von Neumann first visited Los Alamos in September 1943. During that visit he pushed Los Alamos to pursue the implosion method,[4] and to use the PCAM machines for calculations.[14]

Frankel, Nelson, and Namoi Livesay, Figure 10, developed the first program for solving the set of hydrodynamics equations on the PCAM. See the companion ANS article regarding the first hydrodynamics code.[15]

Programming the PCAM started with dividing the equations into IBM operations: add these two quantities, divide these two, sum, etc.[9] This type of programming is shown in detail in LA-1057 (Ref. 13) and LA-1058 (Ref. 16). To illustrate PCAM programming a simple linear interpolation, $c = (a+b)/2$, from section 2.2-2 of LA-1057 (Ref. 13) is described. Linear interpolation was used to fill in the 1,000 cards for the equation-of-state of each material, starting from a sparse grid calculated by hand.

Figure 11 shows a typical 80 column punch card. The card is read by electrical brushes that contact through the holes punched into the card. Interpretation of punch cards is positional, in this case the card is from 1970s with a FORTRAN "print" command in columns 1 through 11, and columns 15 and 16 have the argument to print to unit "40". Cards were manually typed on the 031 keypunch, then retyped on a verifier to catch typing errors. Usually cards generated by the PCAM did not have the human readable label printed across the top of the card, as shown in Figure 11. If this was needed, a card could be feed through the interpreter which would read the card and print the label. As a practical matter, Los Alamos needed the verifier and interpreter listed in Table 1.

To carry out the linear interpolation requires using three PCAM machines, the 513 reproducing punch, the 075 card sorter, and the 601 multiplier. The programmers have to develop a flow sheet for the operators. A pseudo-flow sheet for the linear interpolation is shown in Table 2. There are 11 steps explicitly shown in Table 2, but three more implicit operations are needed to manually move the cards between the reproducer, sorter, and multiplier. But, how does the multiplier know what to do with card 3? Figure 12 shows the plug board for the 601 multiplier that controls the operation of the machine. The programmers have to develop a plug board wiring diagram to tell the operators how to wire the plug board. The wiring diagram for the linear interpolation is shown in Figure 13. The bottom-left connection of "Brushes" 11 to "X-1" shows that if column 11, row 0, on the card is punched, then this plug board operation is activated. The top shows that argument "a" is on columns 25 to 30, and "b" is on columns 35 to 40. The wiring diagram then carries out the rest of the operations shown in Table 2. The operators receive the flow diagram and wiring diagram, wire the machines, and then move the cards between machines as required. This obviously became a non-trivial choreograph for solving the hydrodynamic equations. The IBM PCAM operators who did this are listed in the Appendix.

In March 1944, before the PCAM arrived, the hydrodynamics program was tested on the Marchant calculators with the assistance of Feynman. Each human computer was assigned one operation, passing the partial results from computer to computer in a specified order.[9]

The PCAM started on the first hydrodynamic calculation "within a week of arrival",[4,7] under the direction of Frankel, Nelson, and Livesay. To check the PCAM results Feynman arranged a competition between the computing pool and the PCAM. The computing pool was as fast as the PCAM for couple of days, but tired out after that. The PCAM gave the correct results, and completed the calculation in May 1944 (Ref. 9).

By April 1944 Los Alamos had developed a scientific computing facility for solving partial differential equations. This consisted of the IBM PCAM, the facility for the machines, the staff to operate the machines, the numeric methods for solving the equations, and the staff to setup and interpret the solutions.

## V. FURTHER DEVELOPMENTS

The original IBM 601 could only multiply two numbers, AxB, and could not divide. In May 1944 Johnston suggested that the calculations could be sped up by using a prototype multiplier he had seen at IBM.[7] They "decided to have IBM build triple-product multipliers and machines that could divide", machines that could do AxBxC, and A/B.[7] In early June 1944, Nelson visited an IBM vice president, convincing IBM to build the machines. Los Alamos ordered three triple-product multipliers, two with the divide capability. These machines were delivered in late 1944,[7] and started being used in November 1944. Figure 8 and Figure 9 both show at least one of the IBM 601 modified with a divide unit, as shown in LA-1057 (Ref. 13). With these unique machines, Los Alamos had one of the most powerful computing facilities in the United States.[2]

In September 1944 the IBM Computations were removed from Group T-2, Serber's Diffusion Theory group, to form Group T-6, IBM Computations, led by Nelson and Frankel.[4,6] Nelson continued as the T-6 Group Leader until he left the lab in January 1946.[6]

As Feynman put it, "Mr. Frankel, who started this program, began to suffer from the computer disease".[9] The disease was that the machines were fun to experiment with. As a result, Frankel was moved to Teller's group in F-Division in January 1945.[4,6] Frankel was replaced by Feynman, Figure 14, overseeing the PCAM calculations, with assistance from Metropolis, Figure 15, reporting to Nelson.[7,9] Feynman was acting T-6 Group Leader during April and May 1945 while Nelson was hospitalized for a badly broken leg from a skiing accident.[7]

Feynman was able to motivate the Special Engineering Detachment (SED) staff operating the PCAM by getting permission to tell them for the first time what they were working on.[9] This was probably in March 1945 when no simulations were completed. The SED staff, see the Appendix, then found ways to improve the PCAM efficiency,[9] such that they went from doing less than one simulation per month to 12 problems in the next four months (April to July 1945). A major part of the speedup was taking advantage of the idle machines as a problem ran through a cycle to run up to three problems at a time, using different color cards for each problem.[8, 9]

By late 1944 Los Alamos was trying to understand the radiation from the bombs.[17] However, the Los Alamos PCAM machines were overcommitted, so they reached out via von Neumann to IBM for help.[2, 17] All the PCAM facilities were already committed to war work, so T.J. Watson, CEO of IBM, decided to establish an IBM computing research facility at Columbia University to assist Los Alamos, and to position the company for the coming peace time.[2, 12] This facility, separate from the existing Astronomical Bureau, was the Watson Scientific Computing Laboratory, headed by Wallace Eckert, who returned from the Naval Observatory.[2, 12] Using the Los Alamos procurement priority it was quickly stood up in May 1945 (Ref. 2). Feynman visited and showed the IBM staff how to wire the machines for the Los Alamos calculations.[2] Other visitors included von Neumann, Bethe, Robert Marshak, and Maria and Joe Mayer.[2]

The Watson group solved two time steps per day, and the results were mailed to Los Alamos for analysis.[2] This effort was eventually stopped because it was slower than the analytical methods for solving the radiation equations.

**VI. EPILOGUE**

When Nelson left the lab in January 1946 he was replaced by R.W. Hamming.[18] By December 1946 Bengt Carlson was the group leader for both the Marchant and IBM computations.[1] Carlson invented the discrete-ordinates Sn method to solve the neutron transport equation in 1953 (Ref. 19), which is still extensively used.

The IBM 601 machines were supplemented with newer IBM 602 systems in 1947 and 1948 (Ref. 8). They were all replaced with the new IBM Card-Programmed Electronic Calculator (CPC) 604 machines in 1949.[8] The last CPC was retired in 1956, replaced by the MANIAC and IBM 701 and 704 electronic computers.[8] The punch card machines had quite a run at Los Alamos!

The Los Alamos experience showed the value of numerical computing for science. An example of this was reinforcing the belief in the value of scientific computing by T.J. Watson, the CEO of IBM. Watson established the Watson Scientific Computing Laboratory at Columbia University in 1945 to assist Los Alamos.[2] After the war, the Computing Laboratory quickly surpassed the Los Alamos facility in size and advanced machines, and attracted a large number of scientists in various disciplines.[2] In turn, this lead Watson to direct the building in 1946-1947 of IBM's first electronic computer, the Selective Sequence Electronic Computer (SSEC), which was dedicated to scientific computing. IBM and Los Alamos had a long partnership from the 1940s into the 1960s developing scientific computers.

Probably the most important consequence of the Los Alamos computing facility was the experience gained by the staff. Frankel and Nelson formed the first computer consulting firm in California in 1947, and were heavily involved in starting the small computer industry on the West Coast. Von Neumann was a major figure in starting the large computer industry on the East Coast, defining the von Neumann computer architecture which is still fundamental today. Metropolis led the effort to build a von Neumann machine at Los Alamos, the MANIAC. John Kemeny, one of the SED PCAM operators, was a co-inventor of the BASIC programming language. Richard W. Hamming, who joined Group T-6 in April 1944, went on to Bell Laboratories where he made many contributions to computing, including the first error-correcting-code to fix parity errors.

The scientific managers were also taught the importance of computing. Oppenheimer at the Princeton Institute of Advanced Study was the patron of von Neumann's machine. Teller and von Neumann made sure computing was a core part of the new Livermore laboratory in 1952. William Penney similarly acquired the use of a von Neumann machine for the UK nuclear weapons program, the first Ferranti Mark I computer installed at the University of Manchester in 1951, and acquired a Mark I* in 1954 (Ref. 20). And of course, computing has always been a core part of the Los Alamos weapons program.

During WWII the Los Alamos triple multiple and divide PCAM were the among the most advanced IBM machines. The PCAM were used to calculate neutron diffusion, bombing tables, equation-of-state, and hydrodynamics. This was probably the broadest problem set of any of the PCAM installations during WWII, and foreshadowed the growth of scientific computing soon to be enabled by the electronic computers such as ENIAC and COLOSSUS. The wartime computing at Los Alamos had wide and lasting impact on computing in general by showing that numerical computing could be of great value throughout science. See the companion ANS article on the impact of the Los Alamos computing facility on the broader computing community.[10]

**ACKNOWLEDGMENTS**

I want to thank Nicholas Lewis for many discussions regarding the early computing at Los Alamos. All images are from the National Security Research Center archives of Los Alamos National Laboratory. This work was supported by the US Department of Energy through the Los Alamos National Laboratory. Los Alamos National Laboratory is operated by Triad National Security, LLC for the U.S. Department of Energy NNSA under Contract No. 89233218CNA000001.

**APPENDIX - IBM OPERATORS**

Usually overlooked are the actual operators of the IBM PCAM. Below is a list, probably incomplete, of the IBM operators at Los Alamos during WWII. It is compiled from the author lists of various reports on the IBM problems and LA-1057 (Ref. 13). Not listed below are Namoi Livesay and F. Eleanor Ewing, who supervised the operators. Livesay was hired in February 1944, and Ewing was hired in the summer of 1944 as Livesay's assistant.

Robert Davis
T/5 Jerome Doppelt
Eleanor Ehrlich
J. Elliott
Pfc. Hazel Gensel
T/4 Sam Goldberg
T/4 Alfred Heermans
T/5 Alex Heller
T/4 Daniel Horvitz
T/3 John Johnston
T/4 John Kemeny

T/4 Joseph D. Kington
Harold Ninger
Frances E. Noah
Sgt. Warren Page
Pfc. B.W. Pierson
T/5 Angeline Sniegoski
Pfc Ethel Taylor
T/4 Alan Vorwald
Pvt. Edith Wright
T/3 W. Douglas Zimmerman

Table I. The IBM Punch Card Accounting Machines that together made the Los Alamos computing facility, 1944-1946. All lists agree on the first eight machines, the last two are auxiliary machines whose existence is uncertain[7, 8, 13].

| Machine | Capability |
|---|---|
| Three 601 multipliers originally. Four 601 by November 1944. | Multiply AxB, add, subtract, and print. Three had the unique AxBxC multiply capability, two could also divide. |
| One 405 alphabetic accounting machine or tabulator | Add, subtract, print |
| One 513 reproducing summary punch | Punch output cards, and reproduces batches of cards |
| One 031 alphabetic duplicating punch keypunch | Data entry keyboard that punched cards, and could reproduce a single card |
| One 075 card sorter | Sorting of cards by values of one column |
| One 077 collator | Sorting of cards by values of multiple columns |
| **Possible auxiliary machines** | |
| One 550 alphabetic interpreter (a 552 in 1949) | Copies punched information of one card onto another card and prints a human readable translation on the card |
| One verifier | Retype a card to verify that it was originally typed correctly |

Table 2. PCAM pseudo flow sheet for linear interpolation of c = (a+b)/2.

|  | Operation | Card | Column |
|---|---|---|---|
| **513 Reproducer** | | | |
| 1 | Read a | 1 | 25 - 30 |
| 2 | Read b | 2 | 35 - 40 |
| 3 | Write a, b, 5 | 3 | 25 – 30, 35 – 40, 80 |
| **075 Sorter** | | | |
| 1 | Merge card 3 into card deck that needs c | 3 | N/A |
| **601 Multiplier** | | | |
| 1 | Read card 3, a, b | 3 | 25 – 30, 35 – 40 |
| 2 | Add a+b | N/A | N/A |
| 3 | Read 5 | 3 | 80 |
| 4 | Shift 5 to 0.5 | N/A | N/A |
| 5 | Multiply (a+b)*0.5 | N/A | N/A |
| 6 | Skip to next card | N/A | N/A |
| 7 | Punch c on new card | 4 | 25-30 |

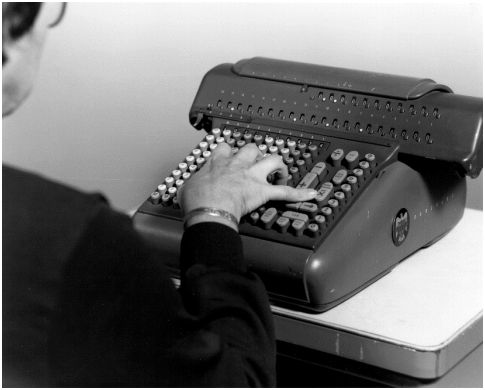

Figure 1. A Marchant calculator being operated by a computer.

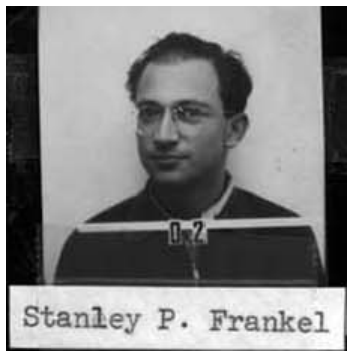

Figure 2. Stanley P. Frankel

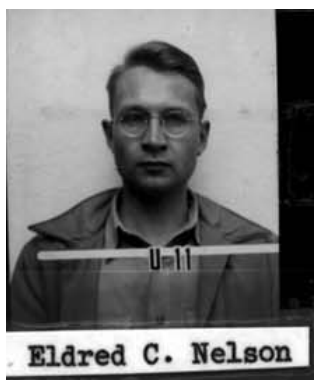

Figure 3. Eldred C. Nelson

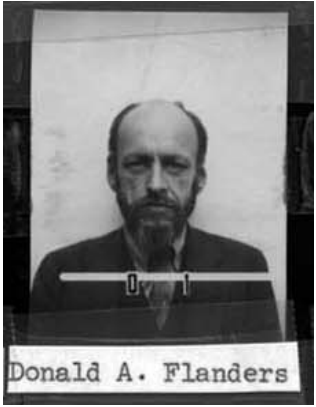

Figure 4. Donald "Moll" Flanders

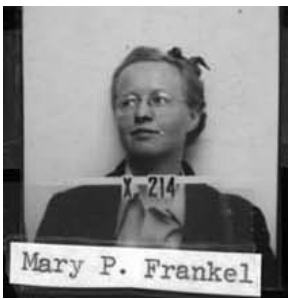

Figure 5. Mary P. Frankel

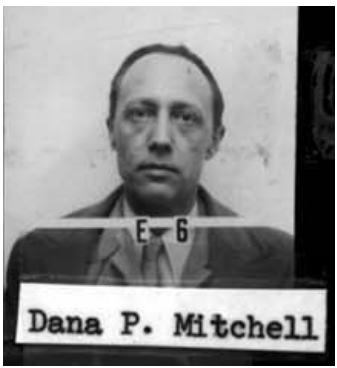

Figure 6. Dana P. Mitchell

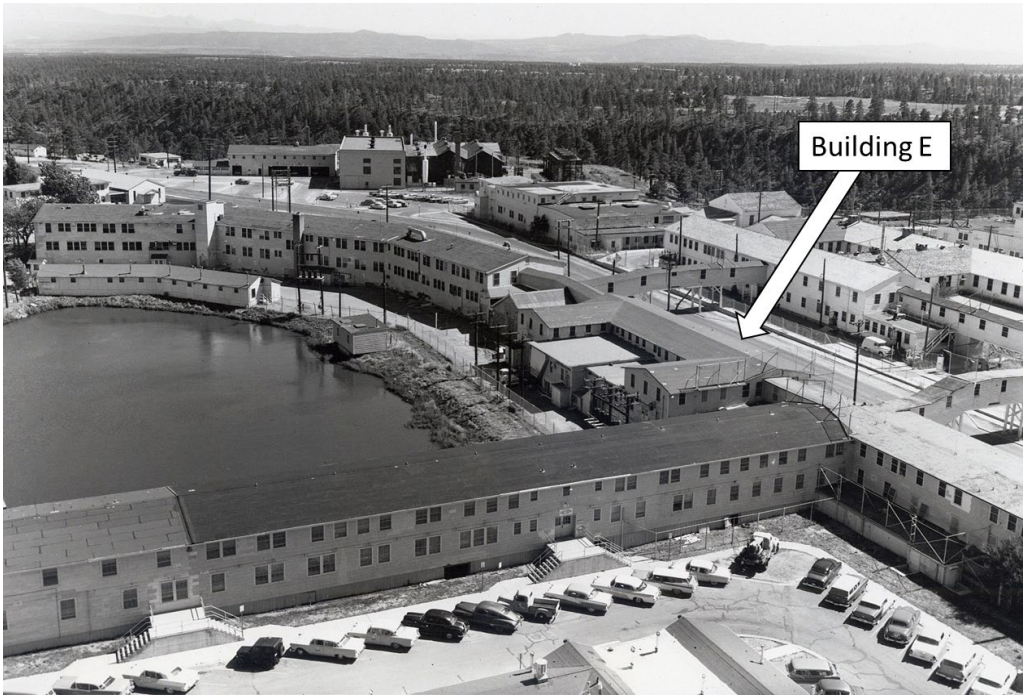

Figure 7. Post-war photo of the Los Alamos Technical Area showing Building E which housed both the hand calculators of Group T-5, and the IBM PCAM of Group T-6.

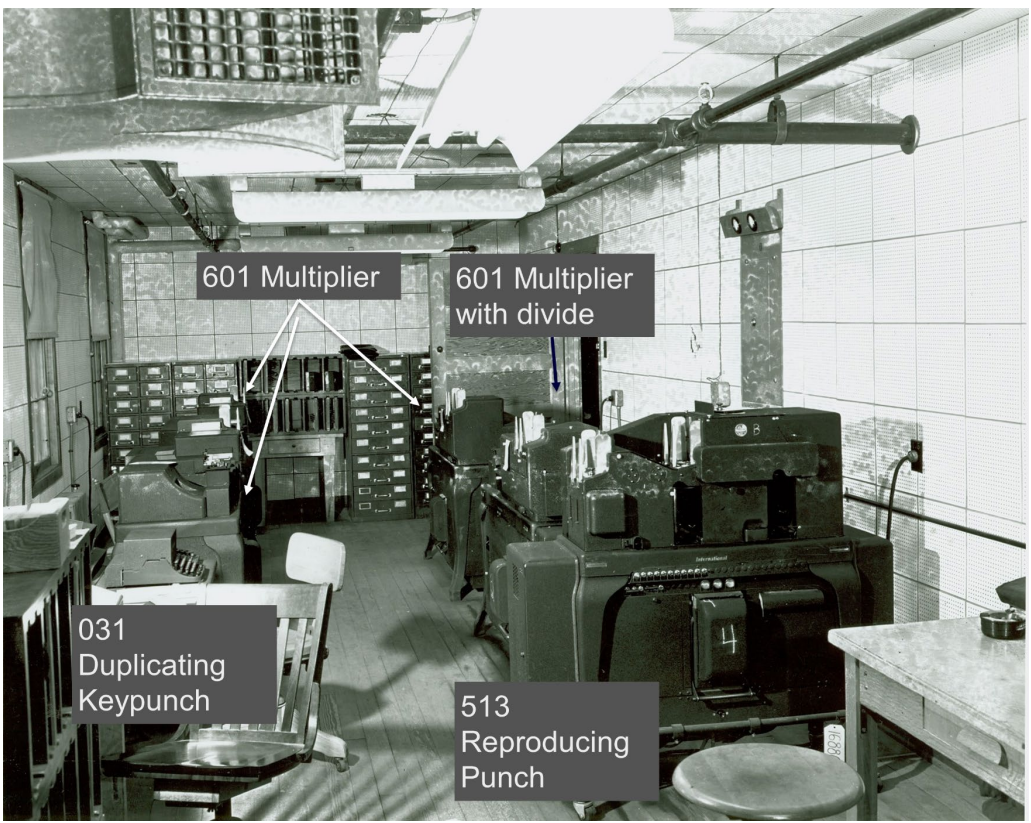

Figure 8. Part of the IBM PCAM installation at Los Alamos during the Manhattan Project, after November 1944. Right side, front to back: 513 reproducer punch, 601 multiplier with divide unit, 601 multiplier without divide unit. Left side, front to back: 031 duplicating keypunch, 601 multiplier, 601 multiplier.

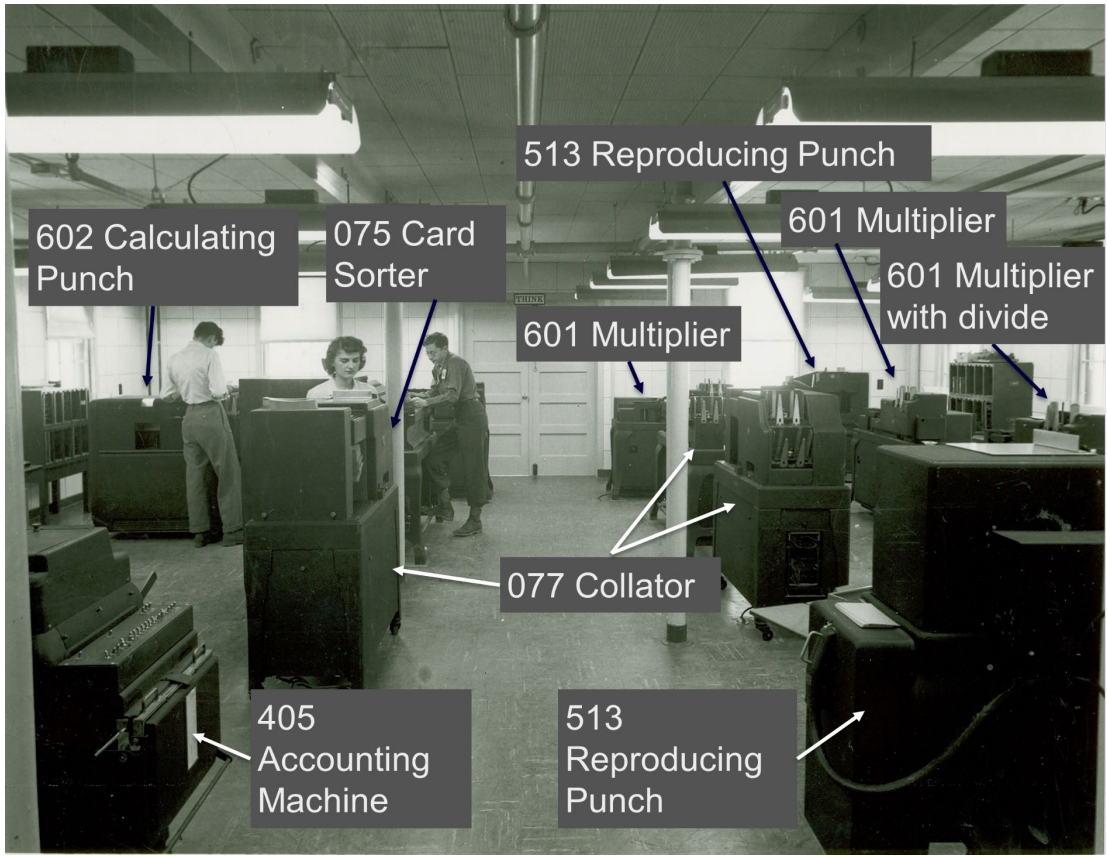

Figure 9. Los Alamos IBM PCAM facility after the first 602 was delivered sometime in 1947. This is the 1945 equipment with the addition of two more 077 collators, one more 513 reproducing punch, and a 602 calculating punch. The 031 keypunch is not shown.

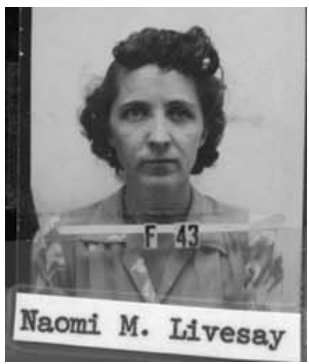

Figure 10. Naomi M. Livesay

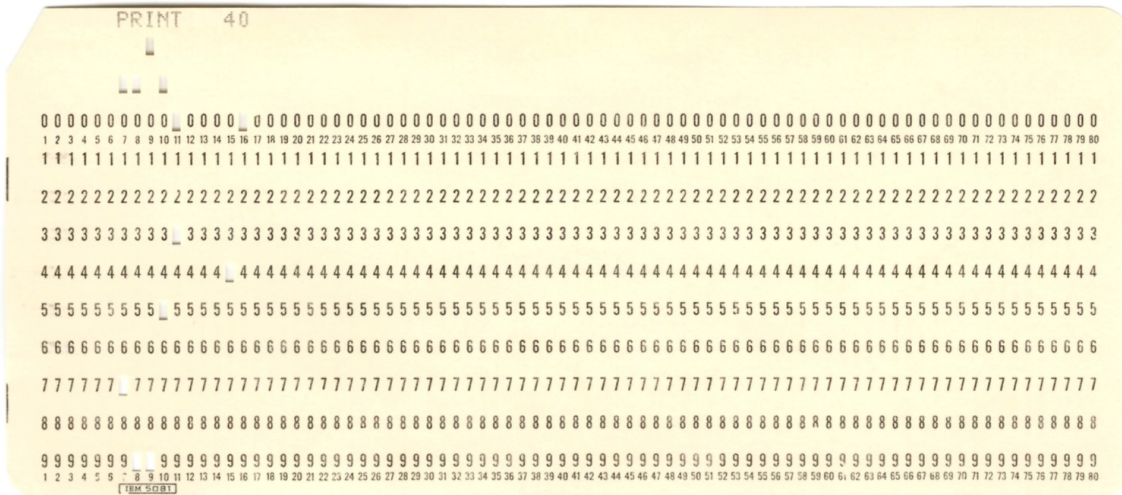

Figure 11. IBM 80-column punch card.

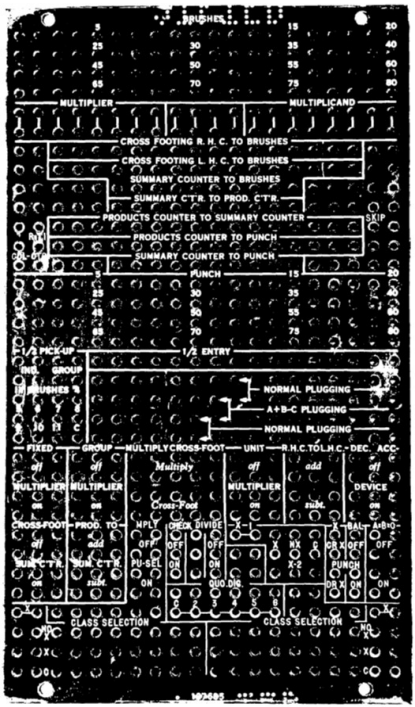

Figure 12. Plug board for the IBM 601 Multiplier.

no

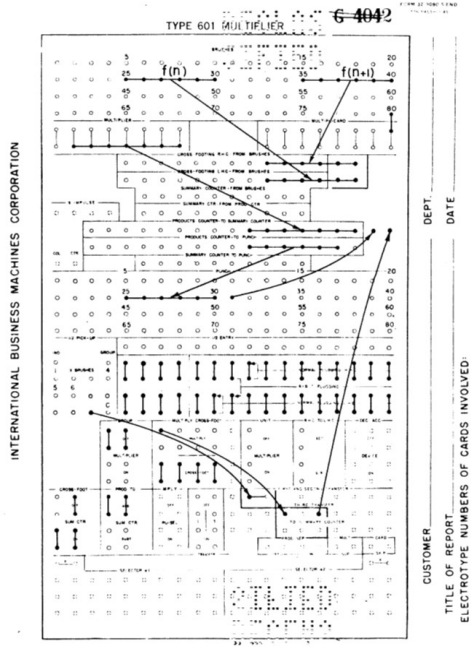

Figure 13. Wiring diagram for a linear interpolation on the IBM 601 Multiplier.

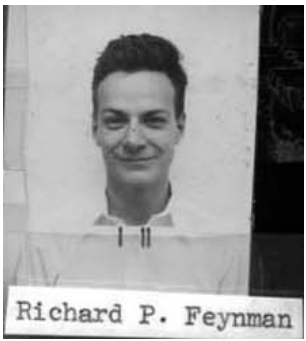

Figure 14. R.P. Feynman

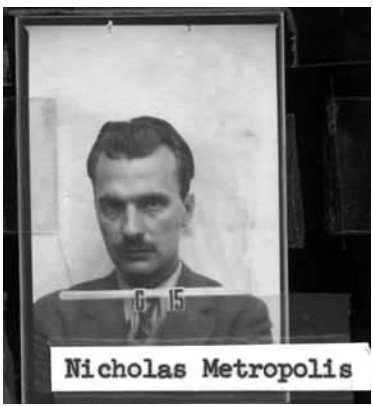

Figure 15. N. Metropolis